\documentclass[12pt]{article}
\newcommand{\be}{\begin{eqnarray}}
\newcommand{\ee}{\end{eqnarray}}
\newcommand{\non}{\nonumber}
\newcommand{\tr}{\mathop{\rm tr}\nolimits}

\begin{document}

\begin{titlepage}
\strut\hfill UMTG--213
\vspace{.5in}
\begin{center}

\LARGE Discrete Symmetries and $S$ Matrix of the XXZ Chain \\[1.0in]
\large Anastasia Doikou and Rafael I. Nepomechie\\[0.8in]
\large Physics Department, P.O. Box 248046, University of Miami\\[0.2in]  
\large Coral Gables, FL 33124 USA\\

\end{center}

\vspace{.5in}

\begin{abstract}
We formulate the notion of parity for the periodic XXZ spin chain 
within the Quantum Inverse Scattering Method.  We also propose an 
expression for the eigenvalues of the charge conjugation operator.  We 
use these discrete symmetries to help classify low-lying $S^{z}=0$ 
states in the critical regime, and we give a direct computation of 
the $S$ matrix.
\end{abstract}

\end{titlepage}

\section{Introduction and summary}

The periodic anisotropic Heisenberg (or ``XXZ'' ) spin chain, with the 
Hamiltonian
\be
H =  {1\over 4} \sum_{n=1}^{N} \left\{
\sigma^{x}_n \sigma^{x}_{n+1}
+ \sigma^{y}_n \sigma^{y}_{n+1} 
+ \Delta \left( \sigma^{z}_n \sigma^{z}_{n+1} - 1 \right) \right\} \,, 
\qquad \vec \sigma_{N+1} \equiv \vec \sigma_{1} \,, 
\label{hamiltonian} 
\ee 
has a long and rich history \cite{bethe} - \cite{kbi}.  It is 
the prototype of all integrable models.  The development of the 
Quantum Inverse Scattering Method (QISM)/algebraic Bethe Ansatz 
\cite{faddeev/takhtajan} systematized earlier results, and paved the 
way for far-reaching generalizations.

Parity symmetry has played a valuable role in continuum quantum field 
theory, including integrable quantum field theory.  (See, e.g., 
\cite{korepin}.) However, the notion of parity for discrete spin 
models, in particular for those which are integrable, has not (to our 
knowledge) been discussed.  We show here that parity has a simple 
realization in the algebraic Bethe Ansatz, involving negation of the 
spectral parameter, i.e., $\lambda \rightarrow -\lambda$.  (See Eq.  
(\ref{parity/second}) below.)

We consider also charge conjugation symmetry \cite{baxter}.  We 
conjecture that Bethe Ansatz states of the XXZ chain with $S^{z}=0$ 
are eigenstates of the charge conjugation operator, with eigenvalues 
$(-1)^{\nu}$, where ${\nu}$ is given by Eq.  (\ref{conjecture}).

Working within the framework of the string hypothesis 
\cite{takahashi/suzuki}, we use these discrete symmetries to help 
classify low-lying $S^{z}=0$ states \cite{jkm}, \cite{woynarovich} in 
the critical regime with $0 < \Delta < 1$.  Moreover, we compute the 
$S$ matrix elements corresponding to these states using the method of 
Korepin\cite{korepin} and Andrei-Destri\cite{andrei/destri}.  Our 
results for the $S$ matrix agree with those obtained by thermodynamic 
methods \cite{babujian/tsvelick}, \cite{kirillov/reshetikhin}.

The outline of this letter is as follows.  In Section 2, after a brief 
review of the algebraic Bethe Ansatz, we define the parity operator, 
and we show how it acts on the fundamental quantities of the QISM 
formalism.  We also review the definition of the charge conjugation 
operator, and we propose an expression for the corresponding 
eigenvalues.  In Section 3 we use the root densities for low-lying 
$S^{z}=0$ states in the critical regime to calculate the parity and 
charge conjugation quantum numbers of these states, as well as the $S$ 
matrix.  We conclude in Section 4 by briefly listing some remaining 
unanswered questions.  A more detailed exposition of these results 
will be given elsewhere \cite{d/n}.

\section{Algebraic Bethe Ansatz and discrete symmetries}

In order to fix notations, we briefly recall the essential elements of 
the algebraic Bethe Ansatz for the XXZ chain.  (See the above-cited 
references for details.) We consider the $A^{(1)}_{1}$ $R$ matrix
\be
R(\lambda)=\left( \begin{array}{cccc}
	a(\lambda)                                  \\
    &         b(\lambda) & c                    \\
	&         c          & b(\lambda)           \\
	&                    &           & a(\lambda)
\end{array} \right) \,, 
\ee 
where
\be
a(\lambda) = {\sinh \left( \mu (\lambda + i) \right)\over \sinh (i \mu)} 
\,, \qquad 
b(\lambda) = {\sinh (\mu \lambda)\over \sinh (i \mu)} 
\,, \qquad   
c = 1 \,.
\ee
We regard $R(\lambda)$ as an operator acting on the tensor product 
space $V \otimes V$, where $V$ is a two-dimensional complex vector 
space. This $R$ matrix is a solution of the Yang-Baxter equation
\be
R_{12}(\lambda-\lambda')\ R_{13}(\lambda)\ R_{23}(\lambda')
= R_{23}(\lambda')\ R_{13}(\lambda)\ R_{12}(\lambda-\lambda') \,,
\label{YB}
\ee 
where $R_{ij}$ are operators on $V \otimes V \otimes V$, with $R_{12} 
= R \otimes 1$, etc.  The $L$ operators 
$L_{0 n}(\lambda) = R_{0 n}(\lambda - {i\over 2})$ act on so-called auxiliary 
($0$) and quantum ($n$) spaces.  
The monodromy matrix $T_{0}(\lambda)$ is defined as a product of $N$ 
such operators
\be
T_{0}(\lambda) = L_{0 N}(\lambda) \cdots L_{0 1}(\lambda) =
\left( \begin{array}{cc}
A(\lambda) & B(\lambda)  \\
C(\lambda) & D(\lambda)
\end{array} \right) 
\,.
\ee
The transfer matrix $t(\lambda)$, defined by tracing over the 
auxiliary space
\be
t(\lambda) = \tr_{0} T_{0}(\lambda) = A(\lambda) + D(\lambda) \,,
\ee
has the commutativity property 
$\left[ t(\lambda)\,, t(\lambda') \right] = 0$. The 
transfer matrix also commutes with the $z$ component of the total 
spin, $S^{z} = {1\over 2} \sum_{n=1}^{N} \sigma^{z}_n$. 
The Hamiltonian
\be
H &=& {i \sin \mu \over 2 \mu} {d\over d \lambda} \log 
t(\lambda)\Big\vert_{\lambda = {i\over 2}} - {N\over 2}\cos \mu  
\ee
coincides with the XXZ Hamiltonian (\ref{hamiltonian}), provided
$\Delta = \cos \mu $.
The critical regime $-1 < \Delta < 1$ corresponds to $\mu$ real, with 
$0 < \mu  < \pi$. The momentum operator $P$ is defined by
$P={1\over i}\log \ t({i\over 2})$,
since $t({i\over 2}) $ is the one-site shift operator.

Let $\omega_{+}= {1 \choose 0}^{\otimes N}$
be the ferromagnetic vacuum vector with all spins up.
The Bethe state $\prod_{\alpha=1}^{M} B(\lambda_{\alpha})\ \omega_{+}$
is an eigenstate of the transfer matrix $t(\lambda)$
if $\{ \lambda_{1} \,, \ldots \,, \lambda_{M} \}$ are distinct and obey 
the Bethe Ansatz equations
\be
\left( {\sinh  \mu \left( \lambda_{\alpha} + {i\over 2} \right) 
\over   \sinh  \mu \left( \lambda_{\alpha} - {i\over 2} \right) } 
\right)^{N} 
= \prod_{\scriptstyle{\beta=1}\atop \scriptstyle{\beta \ne \alpha}}^M 
{\sinh  \mu \left( \lambda_{\alpha} - \lambda_{\beta} + i \right) 
\over 
 \sinh  \mu \left( \lambda_{\alpha} - \lambda_{\beta} - i \right) }
\,, \qquad \alpha = 1 \,, \cdots \,, M \,. 
\label{BAE}
\ee
The corresponding energy and momentum are given by
\be
E = - \sin^{2} \mu  \sum_{\alpha=1}^{M} 
{1\over \cosh (2 \mu \lambda_{\alpha}) - \cos \mu } \,, \qquad   
P = {1\over i} \sum_{\alpha=1}^{M} 
\log {\sinh  \mu \left( \lambda_{\alpha} + {i\over 2} \right) 
\over \sinh  \mu \left( \lambda_{\alpha} - {i\over 2} \right)} 
\quad (\mbox{mod } 2 \pi) \,, 
\ee
and $S^{z} = {N\over 2} - M$. 

\subsection{Parity}

We define the parity operator $\Pi$ on a ring of $N$ spins by
\be
\Pi \ X_{n} \ \Pi^{-1} = X_{N+1-n} \,,
\ee 
where $X_{n}$ is any operator at site $n \in \{ 1 \,, 2\,, \ldots \,, 
N \}$.  Clearly, $\Pi$ acts on the tensor product space $V^{\otimes 
N}$.  We can represent $\Pi$ by
\be
\Pi = \left\{ \begin{array}{cc}
{\cal P}_{1\,, N} {\cal P}_{2\,, N-1} \ldots {\cal P}_{{N\over 2}\,, 
{N+2\over 2}} & \quad \mbox{for} \quad N=\mbox{  even  } \\
{\cal P}_{1\,, N} {\cal P}_{2\,, N-1} \ldots {\cal P}_{{N-1\over 2}\,, 
{N+3\over 2}} & \quad \mbox{for} \quad N=\mbox{  odd  } 
\end{array} \right. \,,
\ee
where ${\cal P}_{i j}$ is the permutation matrix which permutes 
the $i^{th}$ and $j^{th}$ vector spaces. We note that 
$\Pi = \Pi^{-1} = \Pi^{\dagger}$ and hence $\Pi\ \Pi^{\dagger}=1$.

The ``parity invariance'' of the $R$ matrix 
${\cal P}_{12}\ R_{12}(\lambda)\ {\cal P}_{12} =  R_{12}(\lambda)$
implies that the XXZ Hamiltonian is parity invariant, $\Pi \  H \ \Pi = H$. 
Moreover, $\Pi \  S^{z} \ \Pi = S^{z}$ and $\Pi \  P \ \Pi = - P$.

Under parity, the order of the $L$ operators in the 
monodromy matrix is reversed, 
$\Pi\ T_{0}(\lambda)\ \Pi =  L_{0 1}(\lambda) \cdots L_{0 N}(\lambda)$. 
With the help of the ``time-reversal'' invariance of the $R$ matrix
$R_{12}(\lambda)^{t_{1} t_{2}} = R_{12}(\lambda)$ 
($t_{j}$ denotes transposition in the $j^{th}$ space), one can show 
that
\be
\Pi\ T_{0}(\lambda)\ \Pi =  
(-)^{N-1} W_{0}\  T_{0}(-\lambda)^{t_{0}}\  W_{0} \,,
\qquad  W = i \sigma^{y} \,.
\label{result}
\ee
In particular, we see that $\Pi\ t(\lambda)\ \Pi = (-)^{N} t(-\lambda)$.
Evidently, the parity operator does not commute with the transfer 
matrix. We also obtain from (\ref{result}) the fundamental result
\be
\Pi\ B(\lambda)\ \Pi = (-)^{N-1} B(-\lambda) \,.
\label{parity/second}
\ee
We shall use this result, together with the fact 
$\Pi\ \omega_{+} =  \omega_{+}$,
to investigate whether the eigenvectors of the transfer 
matrix are also eigenvectors of the parity operator. 
Since $\{ \Pi \,, P \} = 0$, a Bethe state can be an eigenstate of 
$\Pi$ only if the momentum is $P = 0 \mbox{ or } \pi \quad (\mbox{mod 
} 2 \pi)$.

\subsection{Charge conjugation}

The charge conjugation matrix $C$ is defined (see, e.g., 
\cite{baxter}) by $C= \sigma^{x}$,
since it interchanges the two-component spins ${1 \choose 0}$ and ${0 
\choose 1}$.  We denote by ${\cal C}$ the corresponding operator 
acting on the tensor product space $V^{\otimes N}$,
\be
{\cal C} =  C_{1} \cdots C_{N} \,.
\ee 
It has the properties ${\cal C} = {\cal C}^{-1} = {\cal C}^{\dagger}$,
and hence ${\cal C}\ {\cal C}^{\dagger}=1$.

The invariance of the $R$ matrix under charge conjugation
$C_{1}\ C_{2}\ R_{12}(\lambda)\ C_{1}\ C_{2} = R_{12}(\lambda)$
implies that the monodromy matrix obeys
\be
{\cal C}\ T_{0}(\lambda)\ {\cal C} = C_{0}\ T_{0}(\lambda)\ C_{0} 
\,.
\ee 
In particular, we see that the transfer matrix is invariant under 
charge conjugation ${\cal C}\ t(\lambda)\ {\cal C} = t(\lambda)$,
while the operator $B(\lambda)$ is mapped to $C(\lambda)$,
\be
{\cal C}\ B(\lambda)\ {\cal C} = C(\lambda) \,.
\label{conj/second}
\ee
Moreover, ${\cal C}\ \omega_{+} = \omega_{-}$, where 
$\omega_{-} = {0 \choose 1}^{\otimes N}$ is the ferromagnetic vacuum 
vector with all spins down.

Since $\{ {\cal C} \,, S^{z} \} = 0$, a Bethe state can be an 
eigenstate of ${\cal C}$ only if $S^{z}=0$.  This corresponds to 
$M=N/2$ with $N$ an even integer.

We conjecture that the eigenvectors of the transfer matrix 
with $M=N/2$ are also eigenvectors of ${\cal C}$, with eigenvalues 
$(-)^{\nu}$, where
\be
\nu = {2 i \mu\over \pi} \sum_{\alpha=1}^{N/2} 
\lambda_{\alpha}  + {N\over 2} \qquad (\mbox{mod } 2 ) \,.
\label{conjecture}
\ee 
This conjecture is supported by explicit checks for $N=2$ and $N=4$, 
and it corresponds to the XXZ limit of a result 
\cite{baxter},\cite{jkm} for the XYZ chain.  Unfortunately, Baxter's 
$Q$-operator proof of the XYZ result, which relies on the 
quasi-double-periodicity of certain elliptic functions, does not 
survive the XXZ limit.  (The XYZ result is important, and so it is 
noteworthy that there does not appear to be a proof of it within the 
generalized algebraic Bethe Ansatz.) Certainly, Eq. (\ref{conj/second}) 
alone does not seem to be sufficiently powerful to investigate this 
conjecture.

\section{Low-lying $S^{z}=0$ states}

We now examine some low-lying $S^{z}=0$ states of the critical XXZ 
chain within the framework of the string hypothesis 
\cite{takahashi/suzuki}, \cite{jkm}.
From the so-called root densities, we compute the parity and charge 
conjugation quantum numbers of these states, and we give a 
direct computation of the two-particle $S$ matrix.  

\subsection{Ground state}

For simplicity, we henceforth restrict to the range $0 < \mu < 
{\pi\over 2}$.  The ground state then lies in the sector with $N$ 
even, and is characterized by $M=N/2$ real roots \cite{yang/yang}, 
\cite{takahashi/suzuki}.  A quantity of central importance is the root 
density $\sigma(\lambda)$, which is defined so that the number of 
$\lambda_{\alpha}$ in the interval $[ \lambda \,, \lambda + d\lambda 
]$ is $N \sigma(\lambda) d\lambda$. The root density for the ground 
state is given by $\sigma(\lambda)  = \left( 2 \cosh  \pi \lambda 
\right)^{-1} \equiv  s(\lambda)$.

We now argue that the ground state is a parity eigenstate.  Denoting 
the ground state by $|v \rangle $, we have
\be
|v \rangle = \prod_{\alpha=1}^{N/2} B(\lambda_{\alpha})\ \omega_{+} 
= \exp \left( \sum_{\alpha=1}^{N/2} \log B(\lambda_{\alpha}) \right)\ 
\omega_{+} =
\exp \left( N \int_{-\infty}^{\infty}d\lambda\ \sigma(\lambda) 
\log B(\lambda) \right)\ \omega_{+} \,.
\label{one}
\ee
Moreover, using (\ref{parity/second}), we obtain
\be
\Pi\ |v \rangle &=& (-)^{N/2} \prod_{\alpha=1}^{N/2} B(-\lambda_{\alpha})\ 
\omega_{+} 
= (-)^{N/2} \exp \left( N \int_{-\infty}^{\infty}d\lambda\ \sigma(\lambda) 
\log B(-\lambda) \right)\ \omega_{+} \non  \\
&=& (-)^{N/2} \exp \left( N \int_{-\infty}^{\infty}d\lambda\ \sigma(-\lambda) 
\log B(\lambda) \right)\ \omega_{+} \,,
\label{two}
\ee
where in passing to the last line we have made the change of variables 
$\lambda \rightarrow -\lambda$.  Finally, comparing Eqs.  (\ref{one}) 
and (\ref{two}), and using the fact that the root density 
is an even function $\sigma(-\lambda) = 
\sigma(\lambda)$, we conclude that
\be
\Pi\ |v \rangle = (-)^{N/2} |v \rangle \,.
\label{three}
\ee 

The nondegeneracy of the ground state  implies that this state is also
a charge conjugation eigenstate. The formula (\ref{conjecture}) 
implies that the corresponding eigenvalue is also $(-)^{N/2}$, since
\be
\sum_{\alpha=1}^{N/2} \lambda_{\alpha}
= N \int_{-\infty}^{\infty}d\lambda\ \sigma(\lambda)\ \lambda = 0 \,,
\ee
where again we have made use of the fact that the root density is an 
even function.

\subsection{Two-particle excited states}

There are two distinct two-particle states with $S^{z} = 0$ (again 
with $N$ even and $M=N/2$) \cite{jkm},\cite{woynarovich} which are 
distinguished by their parity and charge conjugation quantum numbers.  
The first state, which we label by (a), is the two-hole state with one 
string of length 2 (i.e., a pair of roots of the form $\lambda_{0} \pm 
{i\over 2}$, with $\lambda_{0}$ real) and all other roots real.  The 
corresponding density is given by
\be
\sigma_{(a)}(\lambda) = s(\lambda) + {1\over N} \left[
\sum_{\alpha=1}^{2} J(\lambda - \tilde\lambda_\alpha) 
- a_{1}(\lambda - \lambda_{0}\,; \mu') \right]
\,, \label{density/(a)}
\ee
where 
\be
J(\lambda) = {1\over 2\pi} \int_{-\infty}^\infty d\omega \ 
e^{-i \omega \lambda}\ {\sinh \left( ({\pi \over \mu}  - 2) 
{\omega \over 2}) \right) \over
2 \sinh \left( ({\pi \over \mu} - 1){\omega \over 2} \right)
\cosh \left( {\omega \over 2} \right)} \,, \quad  
a_1(\lambda\,; \mu) = {\mu \over \pi} 
{\sin \mu \over \cosh(2 \mu \lambda) - \cos \mu} \,,
\ee
and $\mu'$ is the ``renormalized'' anisotropy parameter given by
$\mu' = \pi \mu / (\pi - \mu)$.
Moreover, $\{ \tilde\lambda_{\alpha} \}$  are the hole rapidities, 
and the center of the 2-string is given by $\lambda_{0} = 
(\tilde\lambda_{1} + \tilde\lambda_{2})/2$.

Since the density $\sigma_{(a)}(\lambda) $ is not an even 
function of $\lambda$ for generic values of $\{ \tilde\lambda_{\alpha} 
\}$ , a generalization of the argument (\ref{one}) - (\ref{three}) 
implies that this state is not a parity eigenstate.  However, in the 
``rest frame'' \cite{korepin}
\be
\tilde\lambda_{1} + \tilde\lambda_{2} = 0 \,,
\label{rest}
\ee
the density is an even function, and therefore the state is a parity 
eigenstate, with parity $(-)^{N/2}$. One can verify that in the rest frame 
the momentum is $P = 0 \mbox{  or  } \pi \quad (\mbox{mod } 2 \pi)$, 
which is consistent with the fact $\{ \Pi \,, P \} = 0$.

According to our conjecture, this state is an 
eigenstate of charge conjugation for all values of 
$\{ \tilde\lambda_{\alpha} \}$, with eigenvalue given by the formula 
(\ref{conjecture}). Remarkably, the first term of that formula gives a 
vanishing contribution. We conclude that the charge conjugation 
eigenvalue for this state is also $(-)^{N/2}$.

We consider now the second $S^{z} = 0$ two-particle state, which we 
label by (b).  This is the two-hole state with one ``negative-parity'' 
string of length 1 (i.e., a root of the form $\lambda_{0} + {i \pi 
\over 2 \mu}$, with $\lambda_{0}$ real) \cite{takahashi/suzuki} and 
all other roots real. The density is given by
\be
\sigma_{(b)}(\lambda) = s(\lambda) + {1\over N} \left[
\sum_{\alpha=1}^{2} J(\lambda - \tilde\lambda_\alpha) 
- b_{1}(\lambda - \lambda_{0}\,; \mu') \right]
\,, \label{density/(b)}
\ee
where $b_1(\lambda\,; \mu) = a_1(\lambda + {\pi\over 2 \mu} \,; 
\mu)$, and again $\lambda_{0} = (\tilde\lambda_{1} + \tilde\lambda_{2})/2$.

As in case (a), this state is not a parity eigenstate
for generic values of $\{ \tilde\lambda_{\alpha} \}$. Let us now 
restrict to the rest frame (\ref{rest}). The Bethe vector is 
then given by
$ |v \rangle = B({i \pi \over 2 \mu}) \ 
\prod_{\alpha=1}^{{N\over 2}-1} B(\lambda_{\alpha}) \ \omega_{+} $.
Acting with the parity operator using Eq.  (\ref{parity/second}), we 
obtain
\be
\Pi\ |v \rangle &=& (-)^{N/2} B(-{i \pi \over 2 \mu}) \ 
\prod_{\alpha=1}^{{N\over 2}-1} B(-\lambda_{\alpha}) \ \omega_{+} 
\non  \\ 
 &=& - (-)^{N/2} B({i \pi \over 2 \mu}) \ 
\prod_{\alpha=1}^{{N\over 2}-1} B(-\lambda_{\alpha}) \ \omega_{+} 
\non  \\ 
&=& - (-)^{N/2} |v \rangle \,.
\ee 
In passing to the second line, we have used the quasi-periodicity 
property $B(\lambda \pm {i\pi\over \mu}) = (-)^{N+1}\ B(\lambda)$, 
and to arrive at the last line we use the 
fact that (in the rest frame) the density is an even function of 
$\lambda$.  Thus, the state has parity $-(-)^{N/2}$. 

Unlike case (a), here the first term of the formula (\ref{conjecture}) 
for the charge conjugation eigenvalue does give a contribution, 
namely, $-1$.  We conclude that the charge conjugation eigenvalue for 
this state is also $-(-)^{N/2}$.

We recall that a Boson-antiBoson state with a symmetric 
(antisymmetric) wavefunction has positive (negative) $\Pi$ and ${\cal 
C}$, while for a Fermion-antiFermion state the opposite is true.  
(See, e.g., Refs.  \cite{bjorken/drell}, \cite{klassen/melzer}.) 
Evidently the statistics of the XXZ excitations vary with the 
value of $N$.

\subsection{$S$ matrix}

We define the $S$ matrix $S(\tilde\lambda_1, \tilde\lambda_2)$ by the 
momentum quantization condition \cite{korepin},\cite{andrei/destri}
\be
\left(e^{i p(\tilde\lambda_1) N}\
S(\tilde\lambda_1, \tilde\lambda_2) - 1 \right) 
| \tilde\lambda_1, \tilde\lambda_2 \rangle = 0 \,, 
\label{quantization} 
\ee
where $\tilde\lambda_1$, $\tilde\lambda_2$ are the 
hole rapidities, and $p(\lambda)$ is the hole momentum.
The $S$ matrix eigenvalues are given 
(up to a rapidity-independent phase factor) by
\be
S_{(j)} \sim   
\exp \left\{ i 2\pi N \int_{-\infty}^{\tilde\lambda_{1}}
\left( \sigma_{(j)}(\lambda) - s(\lambda) \right) d\lambda
\right\}
\,, \qquad j = a \,, b \,, 
\label{sim}
\ee 
where $S_{(a)}$ and $S_{(b)}$ are the eigenvalues of the 
$S$ matrix corresponding to states (a) and (b), respectively.

Recalling the expressions (\ref{density/(a)}), (\ref{density/(b)}) for 
the root densities, it is clear that the $S$ matrix eigenvalues have 
the common factor
\be
S_{(0)} &=& \exp \left\{ i 2\pi \sum_{\alpha=1}^{2} 
\int_{-\infty}^{\tilde\lambda_{1}}
J(\lambda - \tilde\lambda_\alpha) \ d\lambda \right\} \non  \\
&=& \exp \left\{ \int_{0}^{\infty} {d\omega\over \omega}
{\sinh \left( ({\pi \over 2\mu'}  - {1\over 2}) \omega \right) 
\sinh \left( i \omega \tilde\lambda \right) \over
\sinh \left( {\pi \omega\over 2\mu'}  \right)
\cosh \left( {\omega \over 2} \right)} \right\}  \non  \\
&=& \prod_{n=0}^{\infty} \Bigl\{ 
{\Gamma \left[ \left( 1 + {\pi\over \mu'} n - i \tilde \lambda 
\right)/2 \right]\over
\Gamma \left[ \left( 1 + {\pi\over \mu'} n + i \tilde \lambda 
\right)/2 \right]}
{\Gamma \left[ \left( 2 + {\pi\over \mu'} n + i \tilde \lambda 
\right)/2 \right]\over
\Gamma \left[ \left( 2 + {\pi\over \mu'} n - i \tilde \lambda 
\right)/2 \right]} \non  \\
& & \times 
{\Gamma \left[ \left( {\pi\over \mu'} ( n + 1 ) + i \tilde \lambda 
\right)/2 \right]\over
\Gamma \left[ \left( {\pi\over \mu'} ( n + 1 ) - i \tilde \lambda 
\right)/2 \right]}
{\Gamma \left[ \left( 1 + {\pi\over \mu'} ( n + 1 )  - i \tilde \lambda 
\right)/2 \right]\over
\Gamma \left[ \left( 1 + {\pi\over \mu'} ( n + 1 )  + i \tilde \lambda 
\right)/2 \right]} \Bigr\} \,,
\label{common}
\ee 
where $\tilde\lambda = \tilde\lambda_{1} - \tilde\lambda_{2}$.
Moreover, we obtain (up to a rapidity-independent phase factor) 
\be
S_{(a)} = S_{(0)}\  
{\sinh \left( \mu' ( \tilde\lambda + i )/2 \right) \over 
 \sinh \left( \mu' ( \tilde\lambda - i )/2 \right) } \,, \qquad 
S_{(b)} = S_{(0)}\  
{\cosh \left( \mu' ( \tilde\lambda + i )/2 \right) \over 
 \cosh \left( \mu' ( \tilde\lambda - i )/2 \right) } \,.
\ee 
This coincides with the $S$ matrix of sine-Gordon/massive Thirring 
model \cite{korepin},\cite{karowski},\cite{zamolodchikov}, provided we 
identify the sine-Gordon coupling constant as
$\beta^{2} = 8 \left( \pi - \mu \right)$.
This result has been obtained for the XXZ chain previously, although 
by less direct means, in Refs.  \cite{babujian/tsvelick}, 
\cite{kirillov/reshetikhin}.  Note that the regime $0 < \mu <
{\pi\over 2}$ in which we work corresponds to the ``repulsive'' 
regime $4\pi < \beta^{2} < 8 \pi$.

\section{Outlook}

A number of issues remain to be explored.  It would be interesting to 
find a proof (or counterexample!) of the formula (\ref{conjecture}) 
for the charge conjugation eigenvalues of the Bethe Ansatz states.  
It may be worthwhile to investigate discrete symmetries in integrable 
chains constructed with higher-rank $R$ matrices, 
\cite{bazhanov},\cite{jimbo} such as $A_{{\mathcal N}-1}^{(1)}$ with 
${\mathcal N} > 2$.  Since these $R$ matrices are not parity 
invariant, neither are the corresponding Hamiltonians.  However, the 
$R$ matrices do have PT symmetry, which may lead to a useful symmetry 
on the space of states.  Moreover, we have not discussed here the 
interesting case of the noncritical ($\Delta > 1$) regime \cite{unpub}.

\section*{Acknowledgments}

We thank F. Essler and V. Korepin for valuable discussions.
This work was supported in part by the National Science Foundation 
under Grant PHY-9870101.

\end{document}